\documentclass{ws-procs9x6}

\begin{document}

\title{Measurement of the Open Charm Cross-Section in 200 GeV Heavy-Ion Collisions at STAR}

\author{Stephen L. Baumgart \footnote{\uppercase{O}n behalf of the \uppercase{STAR C}ollaboration}
}

\address{Yale University, New Haven, CT 06520, USA \\ 
E-mail: stephen.baumgart@yale.edu}

\maketitle

\abstracts{
We report on measurements of open charm production through both hadronic and semi-leptonic decay channels in $\sqrt{s_{NN}}$ = 200 GeV heavy ion collision at
 the STAR experiment at RHIC. We compare experimental results to theoretical predictions from pQCD. The open charm $R_{AA}$ is also presented as evidence of possible medium-induced energy loss.}

\section{Introduction}

Theory predicts that charm quarks are produced by initial gluon fusion during the very early stages of the matter produced in a relativistic heavy-ion collision. These gluon fusion reactions occur through high momentum transfer ($Q^{2}$) processes before the thermalization of the collision fireball [1]. Gluon fusion reactions scale with the number of binary (nucleon-nucleon) collisions, allowing us to hypothesize that the open charm production in STAR will scale with the number of binary collisions at a particular energy.

The charm cross-section can be calculated in $dp_{t}$ slices at the Fixed-Order-next-to-Leading-Log (FONLL) level [2]. This is useful for finding the charm cross-section from non-phontonic electrons out to high momenta. The open charm spectra in this case can be integrated to give a total charm cross-section of $\sigma_{c\overline{c}}^{NLO}$ = $244^{+381}_{-134} \mu b$. In this evaluation, charm is treated as an active flavor. However, when $p_{t} \approx m_{c}$, the uncertainty of the perturbative series becomes large and $\alpha_{s}$ increases quickly [3]. Instead of calculating the charm cross-section in $p_{t}$ slices and then integrating, one can do the calculation entirely in one step. In this case, charm is treated as massive and not an active flavor and gives $\sigma_{c\overline{c}}^{NLO}$ = $301^{+1000}_{-210} \mu b$.  Notice the large systematic error in the positive direction. This stems from the uncertainty in determining the coupling constant, $\alpha_{s}$, at low values of $x$ [3].

Measurements of charm $p_{t}$ spectra allow several possibilities for further study. One possibility is the measurement of the nuclear modification factor, $R_{AA}$, which is the ratio of the A+A cross-section to the p+p cross-section normalized by the number of binary collisions. A decrease in $R_{AA}$ at high $p_{t}$ is generally believed to be a sign of the formation of a Quark Gluon Plasma. This is because the high $p_{t}$ partons will lose energy in the plasma due to induced gluon radiation [4], among other possibilities. Predictions based on kinematic considerations and the dead cone effect show that the high $p_{t}$ suppressions of heavy quarks would be much less than that of light quarks [5]. 

\section{Results}

The Solenoidal Tracker at RHIC (STAR) experiment [6] has been able to measure the charm cross-section through three different methods: direct reconstruction from the hadronic decay channel of the $D^{0}$ meson, evaluation from non-photonic electrons, and a muon measurement.

\subsection{Reconstruction from Hadronic Decays}
The reconstruction of the $D^{0}(\bar{D^{0}}) \rightarrow K^{-}\pi^{+}(K^{+}\pi^{-})$  in 200 GeV Cu+Cu collisions will be discussed here. The $D^{0}$ measurements in d+Au and Au+Au collisions were done using similar procedures. We first use STAR's Time Projection Chamber (TPC) to detect the daughters created from the hadronic decays of $D^{0}$ (From here on $D^{0}$ will be used to mean $\frac{D^{0} + \bar{D^{0}}}{2}$) [7]. The TPC surrounds the interaction region with a $2\pi$ acceptance in the azimuth and $\pm 1.8$ units of pseudorapidity. By correlating the particles' momenta with their mean energy loss (dE/dx) in the TPC gas, various particle species can be identified [7]. Cuts are then applied using this information to maximize the signal/background ratio. 
Once sets of kaon and pion tracks are obtained, their momenta are used to reconstruct the $D^{0}$ invariant mass. 	
Because of the large combinatorial background present in a heavy-ion collisions the $D^{0}$ mass peak is not visible before a background subtraction. Backgrounds are constructed using two different methods. In the first method, one of the daughters of the $D^{0} \rightarrow K\pi$ decay is rotated in momentum space to every 5 degrees between 150 and 210 degrees in the plane transverse to the beam line. The invariant mass spectrum is then reconstructed to generate a background. In the second method, a pion track is combined with a kaon track from a separate but similar collision event in order to reconstruct an invariant mass. Doing this for many possible combinations creates a background mass spectrum. 
\vspace{-0.6cm}
\begin{figure}
\begin{center}
$\begin{array}{c@{\hspace{0.0in}}c}
\vspace{-0.4cm}
\cr{\resizebox{6cm}{!}{\includegraphics{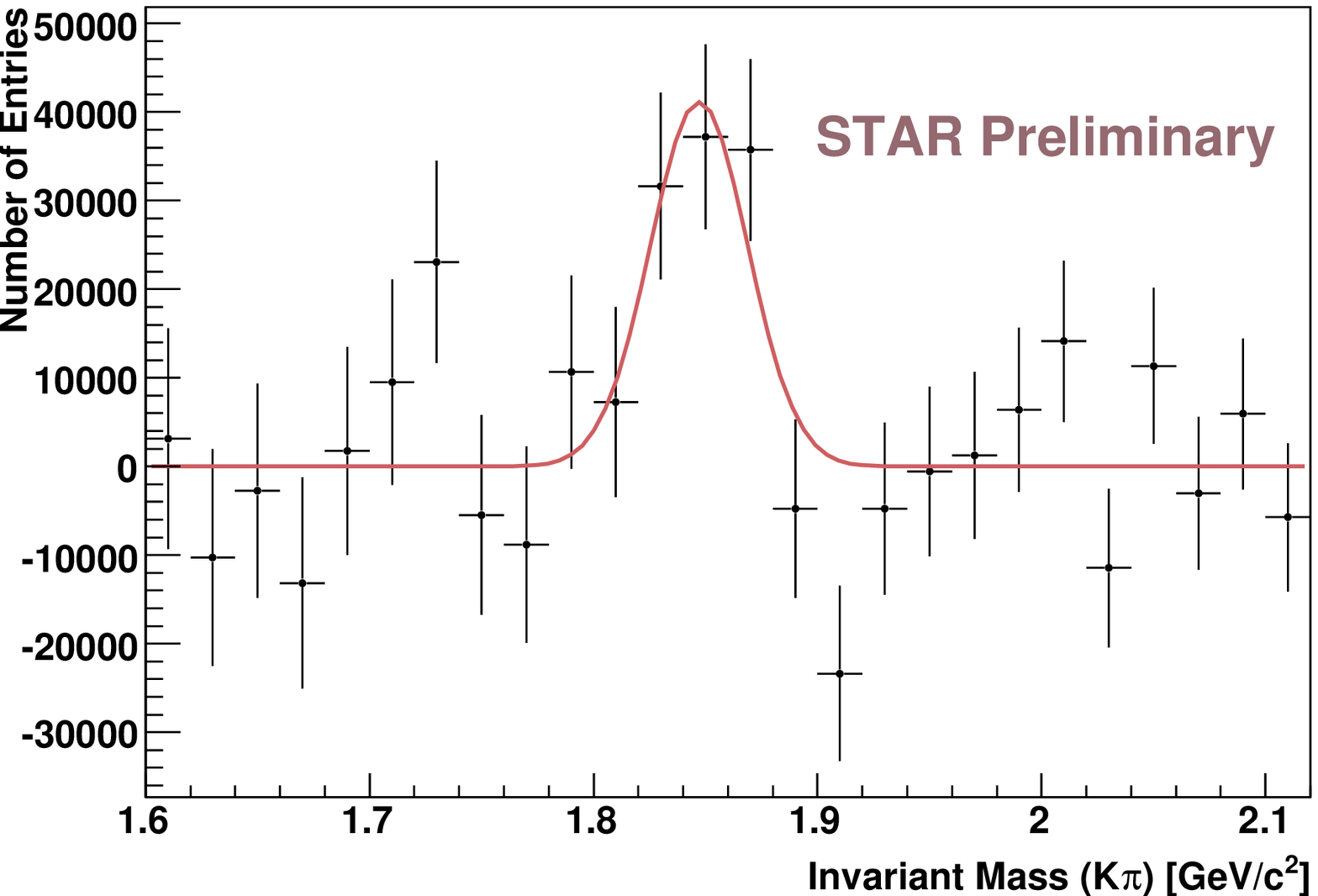}}}
 & {\resizebox{6cm}{!}{\includegraphics{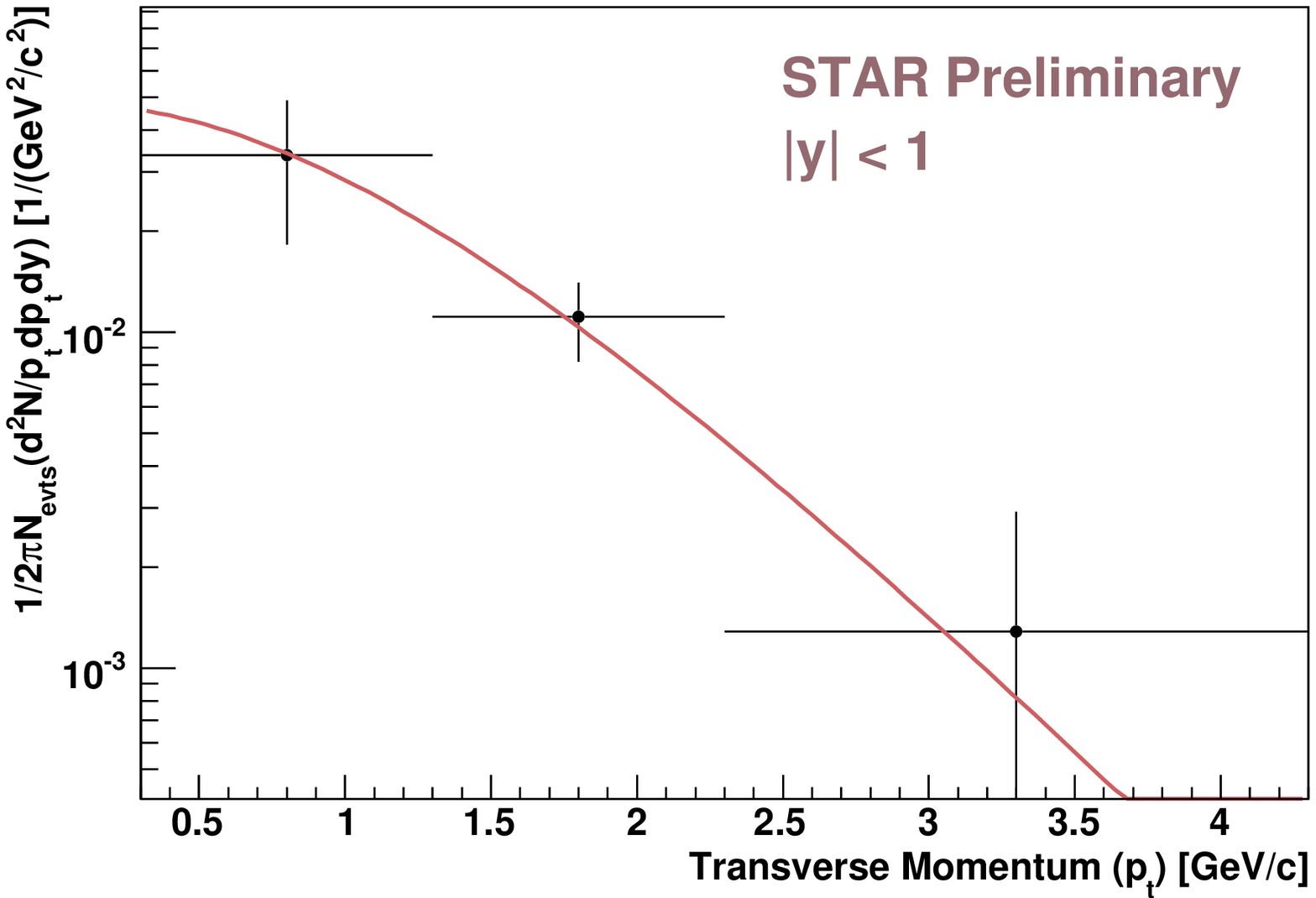}}}
\cr \mbox{\bf } & \mbox{\bf }
\end{array}$
\end{center}
\vspace{-1.0cm}
\caption{Left: The $D^{0} +\bar{D^{0}}$ mass spectrum in 20.2 million $\sqrt{s_{NN}} = 200$ GeV Cu+Cu collisions in the top 60\% centrality after background subtraction. Right: The $(D^{0} +\bar{D^{0}})/2$  spectrum fit with an exponential function in $m_{t} - m_{0}$ .}
\label{D0}
\end{figure}
\vspace{-0.4cm}

After this background is subtracted a residual background remains. We hypothesize that the residual background is created from collective flow effects and from misidentification of resonances. A polynomial function is used to subtract the residual background. We then use a Gaussian function to fit the remaining mass peak. For the full $p_{t}$ range, a peak was found with a significance of 4.3$\sigma$ (See Fig. 1, left). By doing this in $p_{t}$ bins we obtain a spectrum which was then corrected for the branching ratio and for efficiency/acceptance. After integrating an exponential fit to this corrected spectrum we extract a mid-rapidity yield of $\frac{dN}{dy} = 0.36 \pm 0.08$ (stat.) (See Fig. 1, right). Systematic errors are still being evaluated and include the differences between the rotational vs. mixed event background subtractions as well as the limits on the fit to determine residual background.

\subsection{Charm Cross-Section}
The dN/dy yield is extrapolated to the full cross-section for each nucleon-nucleon collision through the following equation,

\vspace{-0.5cm}
\begin{equation}
  \sigma_{c\overline{c}}^{NN} = (dN_{D^{0}}/dy)\times(\sigma_{pp}^{inelastic}/ N_{bin}^{CuCu})\times (f/R)
 \end{equation}

For the Cu+Cu measurement we define $\sigma_{pp}^{inelastic} = 42$ mb as the proton+proton inelastic cross-section [8], $N_{bin}^{CuCu} = 80.4^{+5.9}_{-5.6}$ is the average number of binary collisions in the 60\% most central collisions, $f = 4.7\pm0.7$ is a factor, calculated from simulation, to extrapolate to full rapidity  [9,10], and $R = 0.54 \pm 0.05$ is the ratio of $D^{0}$ to $c\overline{c}$ as measured in $e^{+}e^{-}$ collider experiments [11]. This gives a total charm-cross section of $\sigma_{c\overline{c}}^{NN} = 1.6 \pm 0.3$ mb (stat.)  (See Fig. 3).
\vspace{-1.0cm}
\begin{figure}
\begin{center}
$\begin{array}{c@{\hspace{0.0in}}c}
\vspace{-0.4cm}
\cr{\resizebox{6.5cm}{!}{\includegraphics{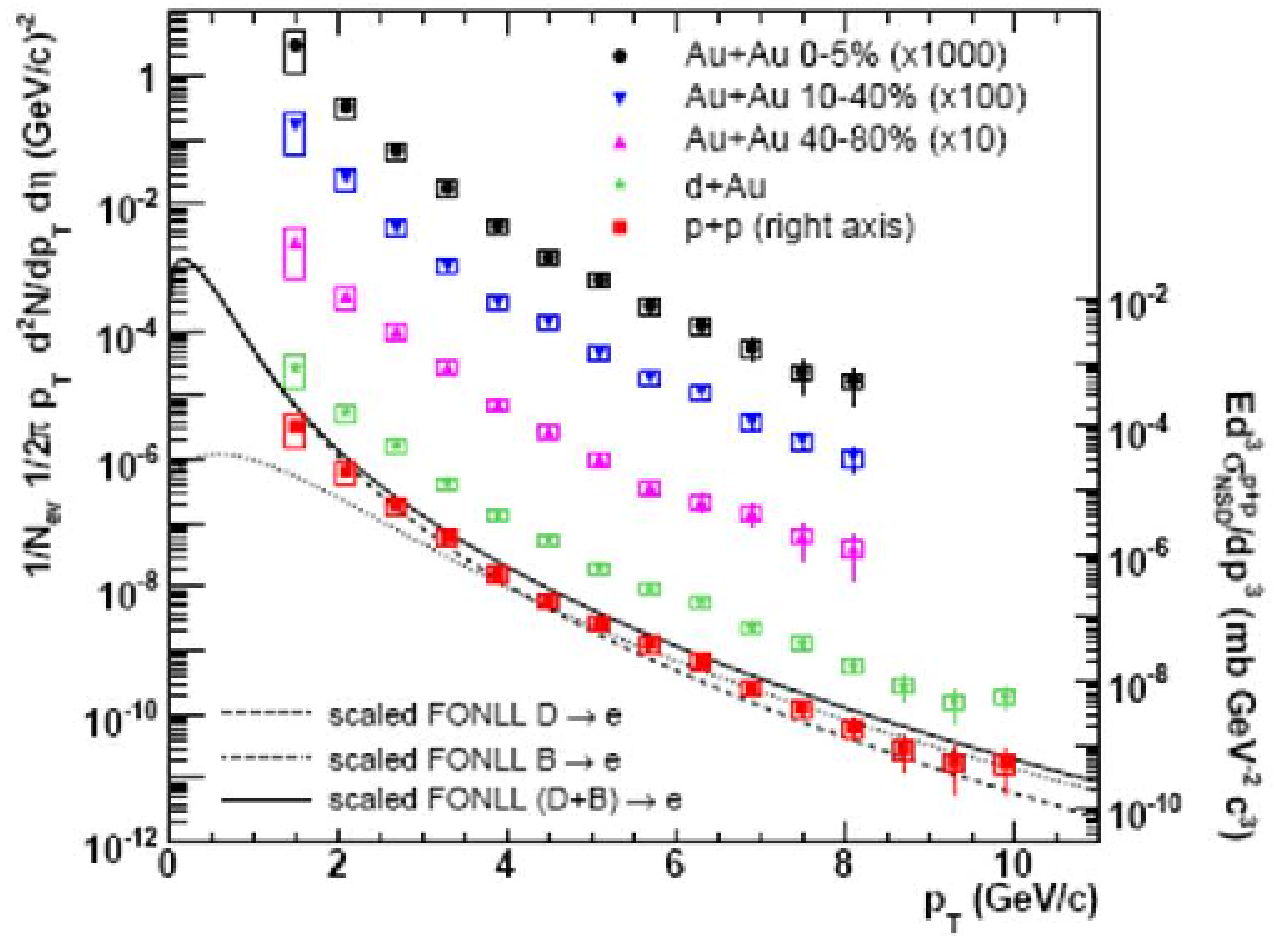}}}
 & {\resizebox{5.0cm}{!}{\includegraphics{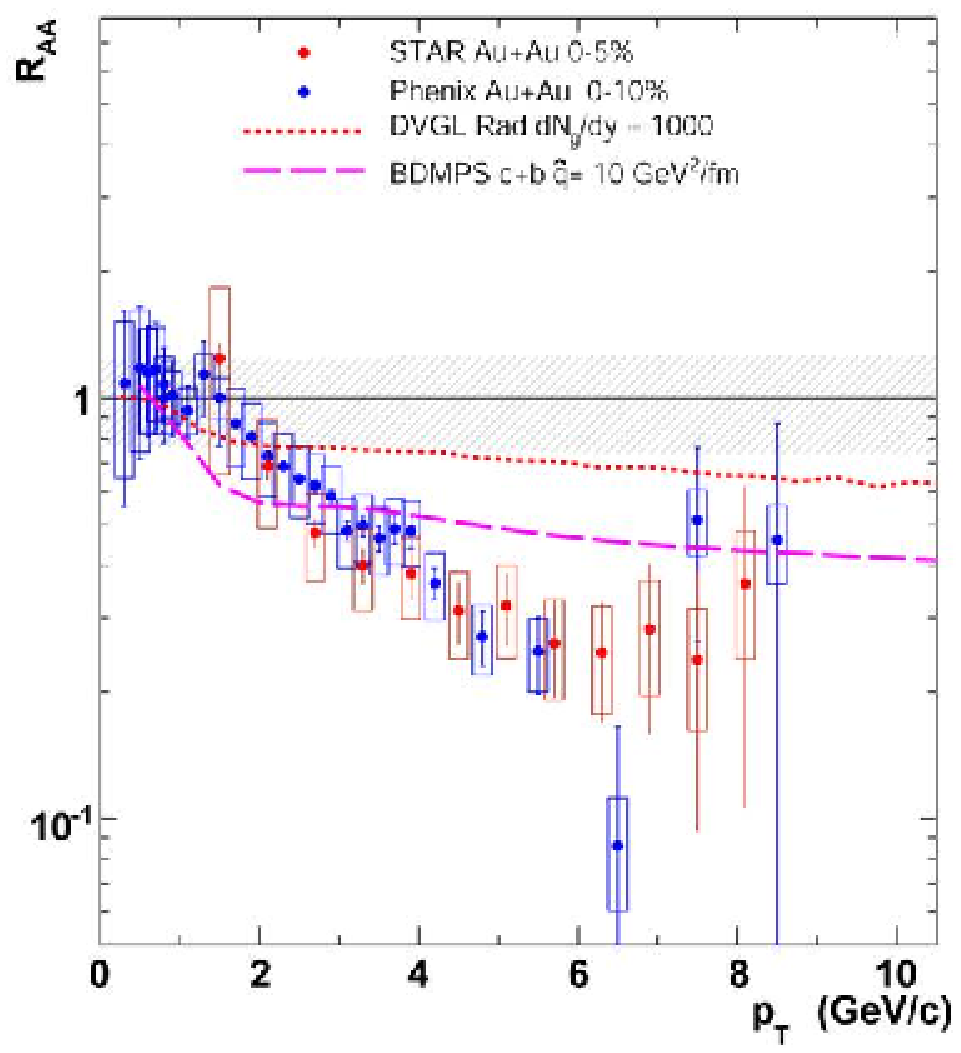}}}
\cr \mbox{\bf } & \mbox{\bf }
\end{array}$
\vspace{-1.0cm}
\end{center}
\caption{Left: The non-photonic electron spectra from multiple systems [8]. The shape of the FONLL curve matches the shape of the data but the yield is higher. Right: The nuclear modification factor $R_{AA}$ from the STAR and PHENIX experiments showing a high-$p_{t}$ suppression of open charm in central Au+Au collisions [13,15,16,17]. } 

\vspace{-0.4cm}
\label{NPE}
\end{figure}
\subsection{Open Charm from Non-Photonic Electrons}
Electrons are identified by their dE/dx energy loss in the TPC as well as their p/E ratios, where the energy is measured by the Barrel Electromagnetic Calorimeter [12, 13]. An invariant mass cut on electron pairs is used to eliminate electrons from photon conversions. The vast majority of the remaining electrons should be the product of heavy flavor decays [14] but a direct hadronic measurement is needed to separate the charm and beauty contributions. The electron spectra are then used to calculate $R_{AA}$. In the 0 to 5\% most central Au+Au collisions STAR has found a suppression of open charm $R_{AA}$ at high $p_{t}$ [13] (See Fig. 2). This suppression can be a sign of medium induced gluon radiation [4]. Surprisingly, this suppression is also as large as that of light quarks [13], contradicting predictions [5]. 


\begin{figure}
\begin{center}
{\resizebox{6.85cm}{!}{\includegraphics{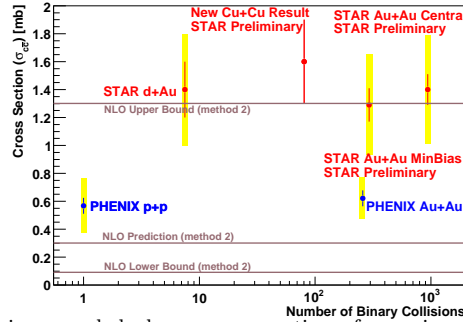}}}
\end{center}
\vspace{-0.6cm}
\caption{The binary-scaled charm cross-sections for various systems measured by STAR and PHENIX and compared to a NLO pQCD prediction. [3, 18, 19, 20] }
\label{Cross Section}
\end{figure}

\section{Interpretation}

The STAR experiment has measured the open charm cross-section using hadrons, non-photonic electrons, and muons. A comparison of STAR open charm results across various systems shows a scaling with the number of binary collisions, consistent with charm production from open gluon fusion (See Fig. 3). These charm cross-sections are near the upper limit of a pQCD NLO predictions and are globally roughly a factor of 2 above PHENIX's. However, PHENIX and STAR's $R_{AA}$ curves match, implying a global normalization error in the electron measurements of one or both experiments. Also, high-$p_{t}$ suppression of open charm in heavy-ion systems points to the influence of a Quark-Gluon Plasma on open charm; however, the suppression is greater relative to that of light quarks than originally predicted.

\end{document}